\documentclass[prd,reprint,twocolumn,notitlepage,preprintnumbers,superscriptaddress]{revtex4}
\usepackage{wrapfig}
\usepackage{verbatim}
\usepackage{amsmath, multirow, amssymb, wasysym, gensymb}
\usepackage{graphicx}
\usepackage{amsfonts}
\usepackage{graphicx}
\usepackage{graphics}
\usepackage[ngerman, english]{babel}
\usepackage{float}
\usepackage{youngtab}
\usepackage{subfigure}
\usepackage[utf8]{inputenc}
\usepackage[usenames,dvipsnames]{xcolor}
\usepackage[normalem]{ulem}
\usepackage{color}

\begin{document}

 \begin{flushright}
 \end{flushright}

\preprint{HEPHY-PUB 965/16}
\preprint{DO-TH 16/09}

\title{Leptoquark patterns unifying neutrino masses,\\ flavor anomalies, and the diphoton excess}
\author{F. F. Deppisch}
\affiliation{Department of Physics and Astronomy, University College London, London WC1E 6BT, United Kingdom}
\author{S. Kulkarni}
\affiliation{Institute of High Energy Physics, Austrian Academy of Sciences, Nikolsdorfergasse 18, 1050 Vienna, Austria}
\author{H. P\"as}
\affiliation{Fakult\"at f\"ur Physik, Technische Universit\"at Dortmund, 44221 Dortmund,
Germany}
\author{E. Schumacher}
\affiliation{Fakult\"at f\"ur Physik, Technische Universit\"at Dortmund, 44221 Dortmund,
Germany}

\begin{abstract} \noindent
Vector leptoquarks provide an elegant solution to a series of anomalies and at the same time generate naturally light neutrino masses through their mixing with the standard model Higgs boson.
We present a simple Froggatt-Nielsen model to accommodate the $B$ physics anomalies $R_K$ and $R_D$, neutrino masses, and the $750$ GeV diphoton excess in one cohesive framework adding only two vector leptoquarks and two singlet scalar fields to the standard model field content. 
\end{abstract}

\maketitle

\section{Introduction}
Over the past years several deviations from the standard model (SM) were reported that thus far remain unresolved. Among the most striking are the rare $B$ decay anomalies that manifest themselves in the ratios 
\begin{align}
R_K = \frac{\text{Br}(B\rightarrow K \mu \mu)}{\text{Br}(B \rightarrow K ee)}\,,\quad R_D=\frac{\text{Br}(B\rightarrow D \tau \nu)}{\text{Br}(B\rightarrow D l \nu)},
\end{align}
with $l=e,\mu$.
The LHCb Collaboration reported a $2.6\,\sigma$ deviation from the SM prediction $R_K^{\text{SM}} = 1.0003 \pm 0.0001$, hinting at a violation of lepton universality. The reported result amounts to  \cite{Aaij:2014ora}
\begin{align}
R_K^{\text{LHCb}} = 0.745 \pm \substack{0.090 \\ 0.074} \pm 0.036\,.
\end{align}
The ratio $R_{D^{(\ast)}}$ has been investigated by several experiments, which all see a slight excess over the SM expectation with a combined statistical significance of more than $4\,\sigma$ \cite{Lees:2013uzd,Huschle:2015rga,Aaij:2015yra},
\begin{align}\begin{array}{llll}
 R_D^{\text{exp}} &= 0.388 \pm 0.047\,, &\quad R_{D^{\ast}}^{\text{exp}} &= 0.321 \pm 0.021\,, \\
 R_D^{\text{SM}} &= 0.300 \pm 0.010\,, &\quad R_{D^{\ast}}^{\text{SM}} &= 0.252 \pm 0.005\,. \\
\end{array}\end{align}

TeV scale leptoquarks modifying $b \rightarrow s ll$ and $b \rightarrow \overline{c} l \nu$ transitions are among the most prominent solutions to the flavor puzzles posed by low-energy precision $B$ physics. Viable candidates to explain the observables $R_K$ and $R_D$ include the scalar leptoquarks $(3,1)_{2/3}$ and $(3,3)_{-1/3}$ \cite{Pas:2015hca,Hiller:2014yaa,Hiller:2014ula,Varzielas:2015iva,Sahoo:2015qha,Sahoo:2015fla,Alonso:2015sja,deBoer:2015boa,Becirevic:2015asa,Gripaios:2014tna,Bauer:2015knc,Sakaki:2013bfa}, denoted by their $SU(3)_C \otimes SU(2)_L \otimes U(1)_Y$ quantum numbers, and their vector equivalents $V_0$ and $V_{1/2}$ \cite{Freytsis:2015qca,Fajfer:2015ycq,Barbieri:2015yvd,Calibbi:2015kma}. Attempts have been made using leptoquarks to draw connections beyond $B$ physics to other unexplained phenomena, such as neutrino masses \cite{Hirsch:1996qy,Helo:2013ika,Pas:2015hca,Mahanta:1999xd,AristizabalSierra:2007nf,Babu:2010vp,Kohda:2012sr,Cai:2014kra,Sierra:2014rxa,Helo:2015fba}, neutrinoless double beta ($0\nu\beta\beta$) decay \cite{Hirsch:1996ye,Pas:2015eia,Helo:2013ika}, $g-2$ \cite{Cheung:2001ip,Baek:2015mea,Bauer:2015knc}, $h\rightarrow \mu\tau$ \cite{Baek:2015mea,Cheung:2015yga}, and even the recently observed diphoton excess near 750 GeV \cite{Murphy:2015kag,Bauer:2015boy}. Vector leptoquarks in particular have been shown to be excellent candidates to explain the latter without the need of introducing many additional fermions  \cite{Murphy:2015kag,deBlas:2015hlv}. For a very recent review on leptoquark physics, see Ref. \cite{Dorsner:2016wpm}.

In this work we propose viable flavor patterns based on a Froggatt-Nielsen (FN) framework for the vector leptoquarks $V_0$ and $V_{1/2}$ to accommodate the $B$ physics anomalies $R_K$ and $R_D$, neutrino masses, and the diphoton excess in one cohesive model. The FN mechanism reproduces the fermion mass hierarchies and quark mixing in excellent agreement with experimental data \cite{Froggatt:1978nt}, while the neutrino-leptoquark interactions give rise to the large leptonic mixing angles.

The paper is structured as follows:
In Sec. \ref{secRKRD} we review effects of the leptoquarks $V_0$ and $V_{1/2}$ on rare $B$ decays to address the observed anomalies in $R_K$ and $R_D$, while accounting for constraints from lepton flavor violation and universality.
Considering the results of Sec. \ref{secRKRD}, we cover possible FN charge assignments to generate the required leptoquark couplings in Sec. \ref{secFN}.
Neutrino mass generation on account of $\Delta L=2$ Higgs-leptoquark mixing is discussed in Sec. \ref{secnu}, while Sec. \ref{secdiph} deals with the vector leptoquark resolution to the 750 GeV diphoton excess. We conclude our study in Sec. \ref{seccon}.

\section{Explaining rare B decays with vector Leptoquarks \label{secRKRD}}
\subsection{$\mathbf{R_K}$}
\begin{figure}[tbh]
\includegraphics[width=0.5\textwidth]{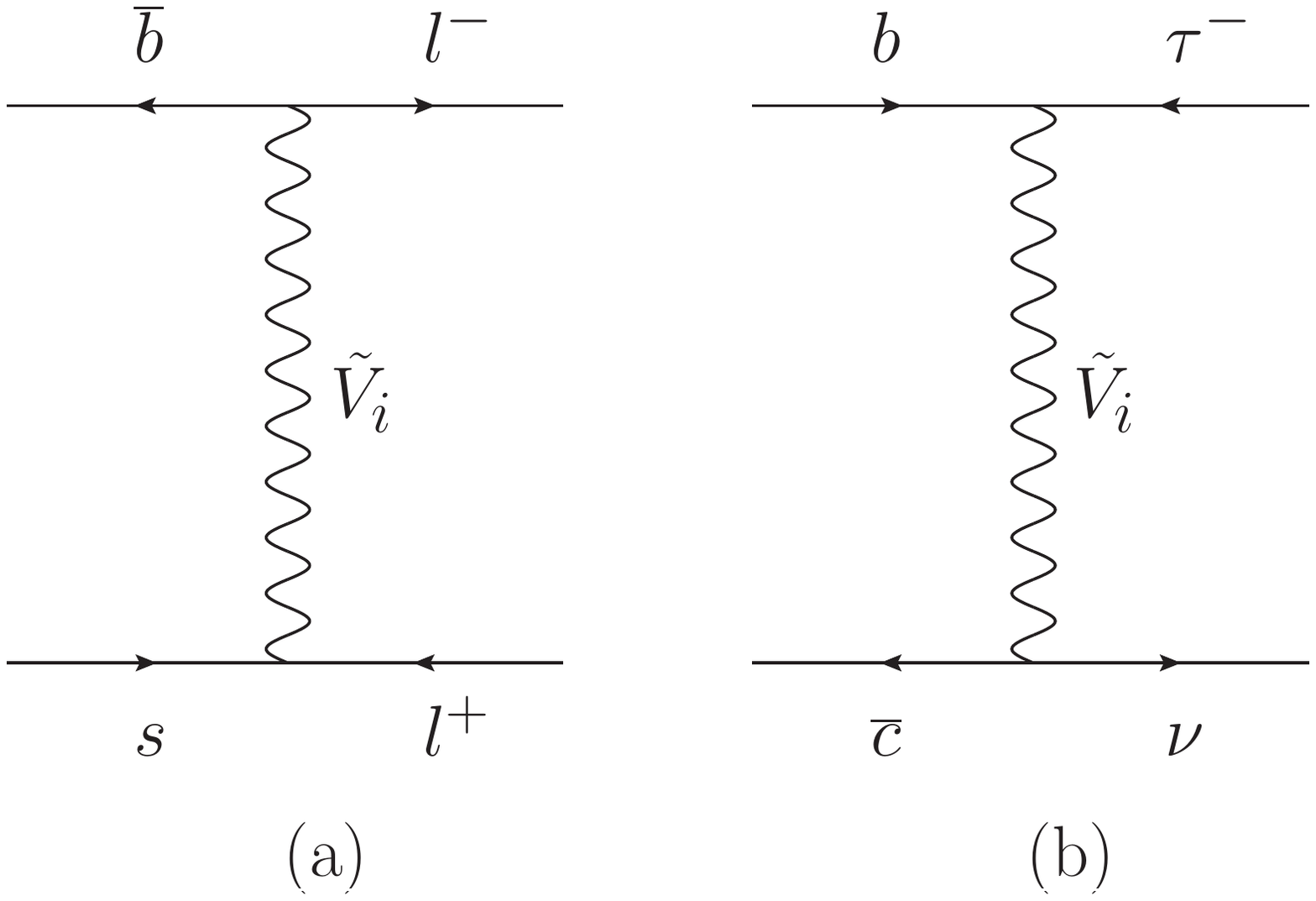}
\vspace{-7cm} 
\caption{(a) $\overline{b} \rightarrow \overline{s}l^+l^-$ transition mediated by vector leptoquarks. $\tilde{V_i}~(i=0,1/2)$ denotes the leptoquark mass eigenstates defined in Eq. (\ref{tanalpha}). (b) Charged current $b \rightarrow \overline{c}\tau\nu$ mediated by $\tilde{V_i}$ modifying the ratio $R_D$.}
\label{bdecays}
\end{figure}

In light of neutrino mass generation we focus only on the vector leptoquarks $V_0$ and $V_{1/2}$ with electric charge $2/3$, which after Fierz rearrangement \cite{Davidson:1993qk} induce (axial) vector operators affecting $B \rightarrow K ll$ as shown in Fig. \ref{bdecays}(a). Their corresponding quantum numbers under the SM symmetries are given in Table \ref{tab:coup}.

 \begin{table}[tbh]
\centering 
\begin{tabular}{|c|c|c|c|c|}\hline
Leptoquark & $(SU(3),SU(2))_{U(1)_Y}$ & $Q_{\text{EM}}$ & $B$ & $L$ \\ \hline
$V_{1/2}$ & $(3,2)_{1/6}$ & $(2/3, -1/3)$ & $1/3$ & $1$ \\ 
$V_{0}$ & $(3,1)_{2/3}$ & $2/3$ & $1/3$ & $-1$ \\ \hline
\end{tabular}
\caption{Quantum numbers of the vector leptoquarks with electric charge $2/3$ that can generate neutrino masses and explain the flavor anomalies.}
\label{tab:coup}
\end{table}

While scalar leptoquarks can be used to combine $R_K$ with neutrino masses \cite{Pas:2015hca}, here we focus on their vector counterparts instead to additionally address $R_D$ and the $750$ GeV diphoton excess recently observed by CMS and ATLAS \cite{ATLAS:2015abc,CMS:2015dxe}. 

To evade tight constraints from low-energy data \cite{Davidson:1993qk} we adopt the typical convention \cite{AristizabalSierra:2007nf} that the leptoquark states $V_0 \equiv V_0^L$ and $V_0^R$, coupling only to left-handed and right-handed fermions, respectively, are independent particles. Of these states we consider only $V_0^L$ for the remainder of this work, as left-handed currents are sufficient to explain the SM deviations.

Recently, a similar analysis based on a $U(2)^5$ flavor symmetry concluded that among the many possible leptoquark mediators, the $(3,1)_{2/3}$ vector leptoquark is the most suitable to explain the anomalies in the $B$ meson sector \cite{Barbieri:2015yvd}. Here we take a different approach to shaping the leptoquark couplings by embedding them into a $U(1)$ FN framework. While Ref. \cite{Barbieri:2015yvd} focused on constraints from the flavor sector, we study in addition how these patterns affect neutrino masses and the diphoton excess.

\vspace{-1cm}

To quantify effects on $R_K$ we work with an effective Hamiltonian
\begin{align}
\mathcal{H}_{\text{eff}} = -4 \frac{G_F}{\sqrt{2}} V_{tb} V_{ts}^{\ast} \frac{\alpha_e}{4 \pi} \sum_i C_i \mathcal{O}_i\,, \label{heff}
\end{align}
where flavor-changing $|\Delta B| = |\Delta S| = 1$ quark transitions are accounted for by the operators $\mathcal{O}_i$ and their Wilson coefficients $C_i$. Furthermore, $G_F$ denotes the Fermi constant, $\alpha_e$ the electromagnetic fine structure constant, and $V_{ud}$ the Cabibbo-Kobayashi-Maskawa (CKM) matrix elements.

After Fierz rearrangement the $V_0$ and $V_{1/2}$ leptoquark couplings
\begin{align}
\label{eq:LLQ}
\mathcal{L}_{\text{LQ}} = \lambda^{L} \overline{Q} \gamma^{\mu} L V_{0,\mu} + \lambda^{R} \overline{u^c} \gamma^{\mu} L V_{1/2,\mu}^{\dagger} + \text{H.c.}\,,
\end{align}
give rise to the effective (axial) vector operators
\begin{align} 
\mathcal{O}_9^l &= \left[\overline{s} \gamma^{\mu} P_L b \right] \left[ \overline{l} \gamma_{\mu} l \right], \\ \mathcal{O}_{10}^l &=  \left[\overline{s} \gamma^{\mu} P_L b \right] \left[ \overline{l} \gamma_{\mu} \gamma_5 l \right]\,.
\end{align}
Note that the $V_{1/2}$ leptoquark shares its quantum numbers with the gauge bosons arising in $SU(5) \rightarrow SU(3)_C \otimes SU(2) \otimes U(1)_Y$ breaking. To avoid rapid proton decay, which is a typical feature of minimal $SU(5)$ models, we assume an underlying symmetry that forbids dangerous diquark operators emerging with $V_{1/2}$.

The leptoquark $V_{1/2}$ itself does not couple directly to down-type quarks, but will do so through its mixing with $V_{0}$. However, as leptoquark mixing is required to be small in order to generate naturally light neutrino masses, any effects on $R_K$ or $R_D$ from $V_{1/2}$ are negligible.

A comparison with Eq. (\ref{heff}) yields $(l=e,\mu)$,
\begin{align}
C_{9}^{l} = - C_{10}^l =  \frac{\pi}{\alpha_e} \frac{{\lambda^{L}_{sl}}^{\ast} \lambda^L_{bl}}{V_{tb} V_{ts}^{\ast}} \frac{\sqrt{2}}{2 m_{V_{0}}^2 G_F}\,, 
\end{align}
where the $ql$ indices denote one element of the matrix $\lambda^{L}$. Consequently, the $R_K$ measurement by LHCb implies  at $1\,\sigma$ \cite{Hiller:2014yaa}
\begin{align}
&0.7 \lesssim~ \text{Re}\left[X^e-X^{\mu}\right] \lesssim 1.5\,, \nonumber \\
\text{with} & \quad X^l = C_9^l - C_{10}^l = 2 C_9^l  \label{xl}\,.
\end{align}
Hence, considering only $V_0$, we obtain
\begin{align} \begin{array}{ll} 
X^e - X^{\mu} &= \displaystyle{\frac{\pi}{\sqrt{2} \alpha_e G_F V_{tb} V_{ts}^{\ast} m_{V_0}^2}} \\ 
&\times \left( \lambda^{L\ast}_{se} \lambda^L_{be} - \lambda^{L\ast}_{s\mu} \lambda^L_{b\mu} \right)\,,
\end{array} \label{xem}
\end{align}
which is equivalent to
\begin{align} \label{rkfin}
\lambda^{L\ast}_{se} \lambda^L_{be} - \lambda^{L\ast}_{s\mu} \lambda^L_{b\mu} \simeq (1.8 \pm 0.7) \cdot 10^{-3} \frac{m_{V_0}^2}{\text{TeV}^2}\,.
\end{align}

It has been shown that right-handed currents lead to deviations in the double ratio $R_{K^{\ast}}/{R_K} \neq 1$ \cite{Hiller:2014ula}, thereby serving as a potential probe of new physics. However, as stated earlier, any possible effects coming from the $V_{1/2}$ leptoquark state are negligible because of the small leptoquark mixing. Therefore, this framework predicts $R_{K^{\ast}}=R_K$.

\subsection{$\mathbf{R_D}$}
While a variety of operators contribute to the tree-level process $b\rightarrow \overline{c}l\nu$ depicted in Fig. \ref{bdecays}(b), several authors pointed out that the vector operator $\mathcal{O}_V$ gives an excellent fit to the $R_{D^{(\ast)}}$ data \cite{Freytsis:2015qca,Fajfer:2015ycq,Calibbi:2015kma},
\begin{align}
\mathcal{O}_V = \left[\overline{c} \gamma^{\mu} P_L b \right] \left[ l \gamma_{\mu} \nu_{l} \right]\,.
\end{align}
In leptoquark UV completions this operator can be provided by both, the $(3,3)_{2/3}$ and $(3,1)_{4/3}$ vector leptoquarks $V_1$ and $V_0$.
The scalar operators $\mathcal{O}_{SL}$ and $\mathcal{O}_{SR}$ can also explain the data but are incompatible with the measured $q^2$ spectra available from BABAR and Belle \cite{Lees:2013uzd,Huschle:2015rga}. This disfavors, e.g, generic two-Higgs-doublet model solutions with a charged scalar contribution.
In our framework the purely left-handed couplings of the leptoquark $V_0$ generate $\mathcal{O}_V$ with the Wilson coefficient
\begin{align}
 C^{cb}_{L,l\nu} = \frac{1}{2\sqrt{2} G_F V_{cb} m_{V_0}^2} \lambda^L_{bl}\lambda_{c\nu}^{L\ast}\,,
\end{align}
which translates to the constraint \cite{Freytsis:2015qca,Fajfer:2015ycq}
\begin{align} \label{rdfin}
 \lambda^L_{b\tau}\lambda^{L\ast}_{c\nu_{\tau}} -  \lambda^L_{b\mu}\lambda^{L\ast}_{c\nu_{\mu}} \simeq (0.18 \pm 0.04) \frac{m_{V_0}^2}{\text{TeV}^2}\,.
\end{align}
Explaining the measurement hence requires a mild hierarchy between the third and the second column of $\lambda^L$ with $\mathcal{O}(1)$ third generation couplings. Furthermore, any explanation of $R_{D^{(\ast)}}$ must also accommodate the SM-like branching ratio of $B \rightarrow \tau \overline{\nu}$ \cite{Lees:2012ju}, requiring further suppression of leptoquark couplings to up quarks.

\subsection{Constraints}
The most stringent constraint that leptoquark models aimed at explaining $R_{D^{(\ast)}}$ have to face, typically comes from nonobservation of the inclusive decay $B \rightarrow X_s \nu\nu$ \cite{Sakaki:2013bfa,Calibbi:2015kma}. As a matter of fact, the $\mathcal{O}(1)$ couplings necessary to explain $R_D$ also affect $b \rightarrow s \nu\nu$ transitions significantly due to $SU(2)$ relations.
What makes $V_0$ such an attractive candidate to explain the rare $B$ decay anomalies is its lack of $\lambda_{d\nu}$ couplings, thereby evading the crucial $B \rightarrow X_s \nu\nu$ constraint.
Nevertheless, lepton-flavor- and universality-violating processes involving down-type quarks and charged leptons are still affected by $V_0$ and have to be taken into consideration. 
Rare kaon decay data places stringent constraints on the first two quark generations \cite{Davidson:1993qk},
\begin{align}
 |\lambda^L_{d\mu} \lambda_{s\mu}^{L\ast} | \lesssim \frac{m_{V_0}^2}{(183\,\text{TeV})^2}\,.
\end{align}
Assuming $m_V \approx 1$ TeV, this implies $|\lambda^L_{d\mu} \lambda_{s\mu}^{L\ast}| \lesssim \epsilon^6$ with $\epsilon \simeq 0.2$.
The couplings required to explain $R_K$ and $R_D$ can also be combined to induce flavor violation. These final states are limited for instance by $B^- \rightarrow K^- \mu \tau$ \cite{Lees:2012zz,Fajfer:2015ycq},
\begin{align}
 |\lambda^L_{b\tau} \lambda^L_{s\mu}| + |\lambda^L_{b\mu} \lambda^L_{s\tau}| \lesssim \epsilon \frac{m_{V_0}^2}{\text{TeV}^2}\,.
\end{align}
On the other hand, constraints from flavor-violating top decays such as $t \rightarrow b \tau \nu_{\tau}$ are rather weak,
\begin{align}
 |\lambda^L_{b\tau} \lambda^L_{t\nu_{\tau}}| \lesssim 4.8 \frac{m_{V_0}^2}{\text{TeV}^2}\,,
\end{align}
as opposed to the flavor-violating lepton decay $\mu \rightarrow e\gamma$, measured by MEG, which constrains \cite{Adam:2013mnn,Varzielas:2015iva}
\begin{align}
 |\lambda_{qe}^L \lambda^L_{q\mu}| \lesssim \frac{m_{V_0}^2}{(34\,\text{TeV})^2}\,.
\end{align}
Thus $|\lambda^L_{qe} \lambda^L_{q\mu}|\lesssim \epsilon^4$ assuming again $m_{V_0} \approx 1$ TeV.

Summarizing the above constraints, an ideal pattern (excluding possible texture-zero solutions) to account for $R_K$ and $R_D$ and to comply with experimental searches would read
\begin{align}
 \lambda^{L} \simeq \left( 
\begin{array}{ccc}
\epsilon^6 & \epsilon^4 & \epsilon^3 \\ 
\epsilon^4 & \epsilon^3 & \epsilon \\ 
\epsilon^3 & \epsilon & 1
\end{array}
\right)\,. \label{idealFN}
\end{align}
The matrix $\lambda_L$ is \textit{a priori} a general matrix, cf. Eq.~(\ref{eq:LLQ}). The symmetric pattern in Eq.~(\ref{idealFN}) is chosen for simplicity while satisfying the experimental constraints. In the following section we study possible $U(1)$ charge assignments to generate such a pattern in a Froggatt-Nielsen framework with two leptoquarks $V_0$ and $V_{1/2}$.

\section{Flavor model}
\label{secFN}
To obtain hierarchical leptoquark patterns as required by low-energy flavor data, one can embed the particle content in an FN framework that not only addresses the SM flavor anomalies, but also explains the fermion mass hierarchies as well as the CKM mixing \cite{Froggatt:1978nt}. Traditionally, the FN mechanism is implemented with a $U(1)$ shaping symmetry and a scalar singlet field $\eta$ charged nontrivially under this $U(1)$. The scalar $\eta$ acquires a vacuum expectation value $v_{\eta}$ at a high scale $\Lambda$, suppressing the nonrenormalizable terms of the Yukawa Lagrangian by a factor $\epsilon^n = \left(\frac{v_{\eta}}{\Lambda}\right)^n \approx 0.2^n$, where $n$ is the sum of the fermion $U(1)_{\text{FN}}$ charges.

Alternatively, one can also employ a discrete $Z_N$ symmetry that in the limit of large $N$ becomes nearly continuous. This avoids further constraints from anomaly cancellation, or extra gauge bosons arising due to the breaking of the continuous gauge symmetry. Model examples that use $Z_N$ symmetries in this manner can be found in Refs. \cite{Hernandez:2016rbi,Hernandez:2015dga,Hernandez:2015hrt,Campos:2014zaa}.

A typical choice of FN charges for the $SU(2)$ doublet fields $\overline{Q}_i$ is $(\overline{Q}_1,\overline{Q}_2,\overline{Q}_3)\,\sim\,(3,2,0)$, which reproduces the quark mixing angles in good agreement with the Wolfenstein parametrization of the CKM matrix. Since the vector leptoquarks $V_{0}$ and $V^{\dagger}_{1/2}$ couple to $\overline{Q}$ and $u$, respectively, their patterns will naturally be hierarchical, unlike their scalar leptoquark counterparts $S^{\dagger}_0$ and $S_{1/2}$ which couple to $Q$ and $\overline{d}$ \cite{Pas:2015hca}.

Evidently, obtaining the ideal pattern given in Eq. (\ref{idealFN}) requires the $\overline{Q}_i$ charges $(3,1,0)$. Such choice of charges, however, leads to a small Cabibbo angle and large mixing among the second and third quark generations contrary to experimental observations. Bearing a little fine-tuning to explain $R_D$ we will therefore focus on the $\overline{Q}_i$ charges $(3,2,0)$. 

Besides CKM mixing, another requirement is that the eigenvalues of the fermion mass matrices reflect the observed hierarchies:
\begin{align}
 \label{e:rel}
 \begin{array}{lllllllllll}
  m_u & : & m_{c} & : & m_{t} & \approx & \epsilon^8 & : & \epsilon^4 & : & 1~,\\
  m_d & : & m_{s} & : & m_{b} & \approx & \epsilon^7 & : & \epsilon^5 & : & \epsilon^3~,\\
  m_e & : & m_{\mu} & : & m_{\tau} & \approx & \epsilon^9 & : & \epsilon^5 & : & \epsilon^3~.\\
 \end{array}
\end{align}
These fermion mass hierarchies fix the $U(1)_{\text{FN}}$ charges of the right-handed quark fields. As yet, in the case of the charged leptons the choice remains ambiguous without any further constraint from mixing.

Finally, the interaction $H i \tau_2 V_{1/2}^{\mu} V_{0\mu}^{\dagger}$ essential for neutrino masses and mixing, dictates the FN charge assignment $Q(V_0) = Q(V_{1/2})$, provided that the Higgs charge $Q(H) = 0$.

The resulting charge assignments can be expressed in terms of the charge $Q(L_3)$, allowing one to suppress the right-handed couplings $\lambda^R$ by choosing different integer values for $Q(L_3) \equiv q_{\tau}$,
\begin{align}
 \lambda^L_{V_0} &\simeq \left( 
\begin{array}{ccc}
\epsilon^6 & \epsilon^4 & \epsilon^3 \\ 
\epsilon^5 & \epsilon^3 & \epsilon^2 \\ 
\epsilon^3 & \epsilon & 1
\end{array}
\right), \label{FNpatt1} \\ \lambda^R_{V_{1/2}} &\simeq \left( 
\begin{array}{ccc}
\epsilon^{8+2q_{\tau}} & \epsilon^{6+2q_{\tau}} & \epsilon^{5+2q_{\tau}} \\ 
\epsilon^{5+2q_{\tau}} & \epsilon^{3+2q_{\tau}} & \epsilon^{2+2q_{\tau}} \\ 
\epsilon^{3+2q_{\tau}} & \epsilon^{1+2q_{\tau}} & \epsilon^{2q_{\tau}}
\end{array}
\right)\,, \label{FNpatt2}
\end{align}
e.g., for $q_{\tau}=5$ we obtain
\begin{align}
 \lambda^R_{V_{1/2}} \simeq \left( 
\begin{array}{ccc}
\epsilon^{18} & \epsilon^{16} & \epsilon^{15} \\ 
\epsilon^{15} & \epsilon^{13} & \epsilon^{12} \\ 
\epsilon^{13} & \epsilon^{11} & \epsilon^{10}
\end{array}
\right)\,. \label{FNpatt3}
\end{align}

\begin{table}[tbh]
\centering 
\begin{tabular}{|c||c|c|c|c|c|c|c|c|c|}\hline
Field & $\overline{Q}_1$ & $\overline{Q}_2$ & $\overline{Q}_3$ & $d$ & $s$ & $b$ & $u$ & $c$ & $t$ \\ \hline
$Q(U(1)_{\text{FN}})$ & 3 & 2 & 0 & 4 & 3 & 3 & 5 & 2 & 0 \\ \hline
\end{tabular}

\vspace{0.2cm}
\begin{tabular}{|c||c|c|c|}\hline
Field & $L_1$ & $L_2$ & $L_3$ \\ \hline
$Q(U(1)_{\text{FN}})$ & $q_{\tau}+3$ & $q_{\tau}+1$ & $q_{\tau}$ \\ \hline
Field & $e$ & $\mu$ & $\tau$ \\ \hline
$Q(U(1)_{\text{FN}})$ & $q_{\tau}-6$ & $q_{\tau} -4$ & $q_{\tau} -3$ \\ 
\hline
\end{tabular}

\vspace{0.2cm}
\begin{tabular}{|c||c|c|c|}\hline
Field & $V_0$ & $V^{\dagger}_{1/2}$ & $H$ \\ \hline
$Q(U(1)_{\text{FN}})$ & $-q_{\tau}$ & $q_{\tau}$ & 0\\ \hline
\end{tabular}
\caption{Possible $U(1)_{\text{FN}}$ quantum numbers to obtain a flavor model with natural fermion mass hierarchies and approximate CKM mixing in good agreement with experimental data. Choosing $q_{\tau}=5$ results in the vector leptoquark patterns discussed in Eq. (\ref{FNpatt3}), while larger values of $q_{\tau} > 5$ will gradually suppress $\lambda^R$ couplings even further.}
\label{tab:FNup}
\end{table}

The FN charges of Table \ref{tab:FNup} yield the following fermion mass matrices up to $\mathcal{O}(1)$ coefficients
\begin{align}
 &M_u \simeq  \left( 
\begin{array}{ccc}
\epsilon^8 & \epsilon^5 & \epsilon^3 \\ 
\epsilon^7 & \epsilon^4 & \epsilon^2 \\ 
\epsilon^5 & \epsilon^2 & 1
\end{array}
\right),\quad M_d \simeq \left( 
\begin{array}{ccc}
\epsilon^{7} & \epsilon^{6} & \epsilon^{6} \\ 
\epsilon^{6} & \epsilon^{5} & \epsilon^{5} \\ 
\epsilon^{4} & \epsilon^{3} & \epsilon^{3}
\end{array}
\right), \nonumber \\
&\qquad \quad M_l \simeq \left( 
\begin{array}{ccc}
\epsilon^{9} & \epsilon^{7} & \epsilon^{6} \\ 
\epsilon^{7} & \epsilon^{5} & \epsilon^{4} \\ 
\epsilon^{6} & \epsilon^{4} & \epsilon^{3}
\end{array}
\right)\,. \label{ml}
\end{align}

The fermion mixing matrices that are required to rotate $\lambda^{L,R}$ into the mass basis follow directly from Table \ref{tab:FNup} and are approximately given by
\begin{align} \label{e:mix}
 V_{u,d}^L &\simeq  \left( 
\begin{array}{ccc}
1 & \epsilon & \epsilon^3 \\ 
\epsilon & 1 & \epsilon^2 \\ 
\epsilon^3 & \epsilon^2 & 1
\end{array}
\right),\quad V_l^{L,R} \simeq \left( 
\begin{array}{ccc}
1 & \epsilon^{2} & \epsilon^{3} \\ 
\epsilon^{2} & 1 & \epsilon \\ 
\epsilon^{3} & \epsilon & 1
\end{array}
\right), \\ \label{e:mix2}
\quad V_u^R &\simeq \left( 
\begin{array}{ccc}
1 & \epsilon^{3} & \epsilon^{5} \\ 
\epsilon^{3} & 1 & \epsilon^{2} \\ 
\epsilon^{5} & \epsilon^{2} & 1
\end{array}
\right),\quad V_d^{R} \simeq \left( 
\begin{array}{ccc}
1 & \epsilon & \epsilon \\ 
\epsilon & 1 & 1 \\ 
\epsilon & 1 & 1
\end{array}
\right)\,.
\end{align}
Although the mixing between the second and third lepton generations is enhanced, the FN mechanism is not feasible to explain the large Pontecorvo–Maki–Nakagawa–Sakata (PMNS) mixing angles. Instead, the fundamental difference between the hierarchical CKM and the anarchical PMNS matrix is attributed to neutrino-leptoquark interactions.

The patterns $\lambda^{L,R}$ have to be rotated into their respective mass bases to account for CKM and PMNS mixing. These new matrices are defined as follows:
\begin{align} \begin{array}{ll}
 \tilde{\lambda}^L_{dl} = V_d^L \lambda^L_{V_0} V_l^{L\dagger}, \qquad &\tilde{\lambda}^L_{u\nu} = V_u^L \lambda^L_{V_0} V_{\nu}^{L\dagger}\,, \\
 \tilde{\lambda}^R_{ul} = V_u^R \lambda^R_{V_{1/2}} V_l^{L\dagger}, \qquad &\tilde{\lambda}^R_{u\nu} = V_u^R \lambda^R_{V_{1/2}} V_{\nu}^{L\dagger}\,.
\end{array} \end{align}
Since all relevant mixing matrices of Eqs. (\ref{e:mix})--(\ref{e:mix2}) are approximately diagonal, the general structure of the leptoquark patterns $\lambda^{L,R}$ remains unchanged when rotating from the symmetry into the fermion mass basis. We can henceforth assume $\tilde{\lambda}^{L,R} \simeq \lambda^{L,R}$, with one exception being $\tilde{\lambda}^L_{u\nu}$ that receives large mixing from $V_{\nu}^L$. The magnitudes of the mixing parameters in $V_{\nu}^L$ can be derived from the experimentally observed PMNS mixing matrix combined with our predictions for $V_l^L$. From
\begin{align}
 U_{\text{PMNS}} = V_l^{L\dagger} V_{\nu}^L\quad \Leftrightarrow \quad V_{\nu}^{L\dagger} =  U_{\text{PMNS}}^{\dagger} V_l^{L\dagger}
 \end{align}
 we infer
 \begin{align}
V_{\nu}^{L\dagger} \sim \left( 
\begin{array}{ccc}
1 & 1 & \epsilon \\ 
1 & 1 & 1 \\ 
\epsilon & 1 & 1
\end{array}
\right),\, \quad \tilde{\lambda}^L_{u\nu} \simeq \left( 
\begin{array}{ccc}
\epsilon^4 & \epsilon^3 & \epsilon^3 \\ 
\epsilon^3 & \epsilon^2 & \epsilon^2 \\ 
\epsilon & 1 & 1
\end{array}
\right)\,.
\end{align}
All of the obtained patterns are valid only up to $\mathcal{O}(1)$ coefficients, allowing us to estimate the extent of tuning required to accommodate the observables $R_K$ and $R_D$. Including the $\mathcal{O}(1)$ coefficients, the relevant coupling matrices read
\begin{align}
 \tilde{\lambda}^L_{dl} &= \left( 
\begin{array}{ccc}
a_{de} \epsilon^6 & a_{d\mu} \epsilon^4 & a_{d\tau} \epsilon^3 \\ 
a_{se} \epsilon^5 & a_{s\mu} \epsilon^3 & a_{s\tau} \epsilon^2 \\ 
a_{be} \epsilon^3 & a_{b\mu} \epsilon & a_{b\tau} 
\end{array}
\right)\,, \label{FNexact} \\ \tilde{\lambda}^L_{u\nu} &= \left( 
\begin{array}{ccc}
a_{ue} \epsilon^4 & a_{u\mu} \epsilon^3 & a_{u\tau} \epsilon^3 \\ 
a_{ce} \epsilon^3 & a_{c\mu}\epsilon^2 & a_{c\tau} \epsilon^2 \\ 
a_{te} \epsilon & a_{t\mu} & a_{t\tau}
\end{array}
\right)\,. \label{FNexact2}
\end{align}
From Eq. (\ref{rkfin}) we get
\begin{align}
 a_{b\mu}^{\ast} a_{s\mu} \simeq -(1.1 \pm 0.4)\frac{m_{V_0}^2}{\text{TeV}^2},
\end{align}
which is a perfect match with $R_K$ data for $m_{V_0} \approx 1$ TeV. On the other hand, the measurement of $R_D$ demands (Eq. (\ref{rdfin}))
\begin{align}
 a_{b\tau} a_{c\tau}^{\ast} - 0.2 \cdot a_{b\mu} a_{c\mu}^{\ast} \simeq (4.5 \pm 1.0) \frac{m_{V_0}^2}{\text{TeV}^2}\,,
\end{align}
which requires a little more fine-tuning that can be accommodated easily with couplings mildly larger than $1$.

Since the $\lambda_{u\nu}^L$ couplings are slightly enhanced due to the large neutrino mixing, it is suggestive to study up-type flavor transitions to make predictions for $D$ meson decay channels with dineutrino final states. Charm constraints are relatively weak compared to those from the kaon sector, cf. Ref. \cite{deBoer:2015boa}. 


In our framework, the most promising channel to search beyond the SM physics is $D^+ \rightarrow \pi^+ \nu \nu$, governed by the couplings $|\tilde{\lambda}^L_{c\nu}\tilde{\lambda}^L_{u\nu}| \approx \epsilon^5$, while predictions for other channels involving charged lepton final states suffer more severe suppression to comply with $K$ physics.

\section{Generating neutrino masses \label{secnu}}
As shown in Refs. \cite{Hirsch:1996qy,Kosnik:2012dj}, two leptoquarks sharing the same electric charge $Q$ will eventually mix through a coupling with the SM Higgs boson via
\begin{align}
V(V_i,H) =&\,h_{V} H i \tau_2 V_{1/2}^{\mu} V_{0\mu}^{\dagger} \nonumber \\ -& (m_{V_i}^2 - g_{V_i} H^{\dagger} H) V_{i,\mu}^{\dagger} V_i^{\mu}\,.
\end{align}
The first term, in particular, accounts for the mixing and hence induces neutrino masses if $h_V~\neq~0$.

The resulting leptoquark mass eigenstates are a mixture of flavor states with 
$Q_{\text{EM}}$ charge $2/3$ and a distinct $-1/3$ state stemming from $V_{1/2}$,
\begin{align}\begin{array}{ll}
&M^2_{2/3} = \left( 
\begin{array}{cc}
m_{V_0}^2 - g_{V_0} v_{\text{SM}}^2  & h_V v_{\text{SM}}\\
h_V v_{\text{SM}} & m_{V_{1/2}}^2 - g_{V_{1/2}} v_{\text{SM}}^2
\end{array}
\right), \\
&M^2_{-1/3} = m_{V_{1/2}}^2 - g_{V_{1/2}} v_{\text{SM}}^2\,.                                                                                                                                                                                                                                                      \end{array}                              \end{align}
The rotation angle $\alpha$ diagonalizing the $M_{2/3}^2$ matrix is determined by
\begin{align}
 \left( 
\begin{array}{c}
\tilde{V}_0 \\ 
\tilde{V}_{1/2}%
\end{array}%
\right) &= R \left( 
\begin{array}{c}
V_0 \\ 
V_{1/2}%
\end{array}%
\right), ~R = \left( 
\begin{array}{cc}
\cos \alpha  & -\sin \alpha\\
\sin \alpha & \cos \alpha
\end{array}
\right) \nonumber \\
 \text{with} \quad &\tan 2\alpha = \frac{2 h_V v_{\text{SM}}}{m_{V_{1/2}}^2 - m_{V_{0}}^2} = \frac{2 h_V v_{\text{SM}}}{\Delta m_V^2}\,, \label{tanalpha}
\end{align}
where $\tilde{V_i}$ denotes the leptoquark mass eigenstates. 

The dimensionful parameter $h_V$ cannot be arbitrarily large, but is in fact limited by the condition of positive leptoquark masses and the perturbativity of the theory to
\begin{align}
 &h_V \leq m_{V_0}^{\prime} m_{V_{1/2}}^{\prime} / v_{\text{SM}} \nonumber \\
 &\text{with}\quad  m_{V_i}^{\prime} \equiv \sqrt{m_{V_i}^2-g_{V_i} v_{\text{SM}}^2}\,.
\end{align}

\begin{figure}[tbh]
\vspace{-1cm}
\includegraphics[width=0.5\textwidth]{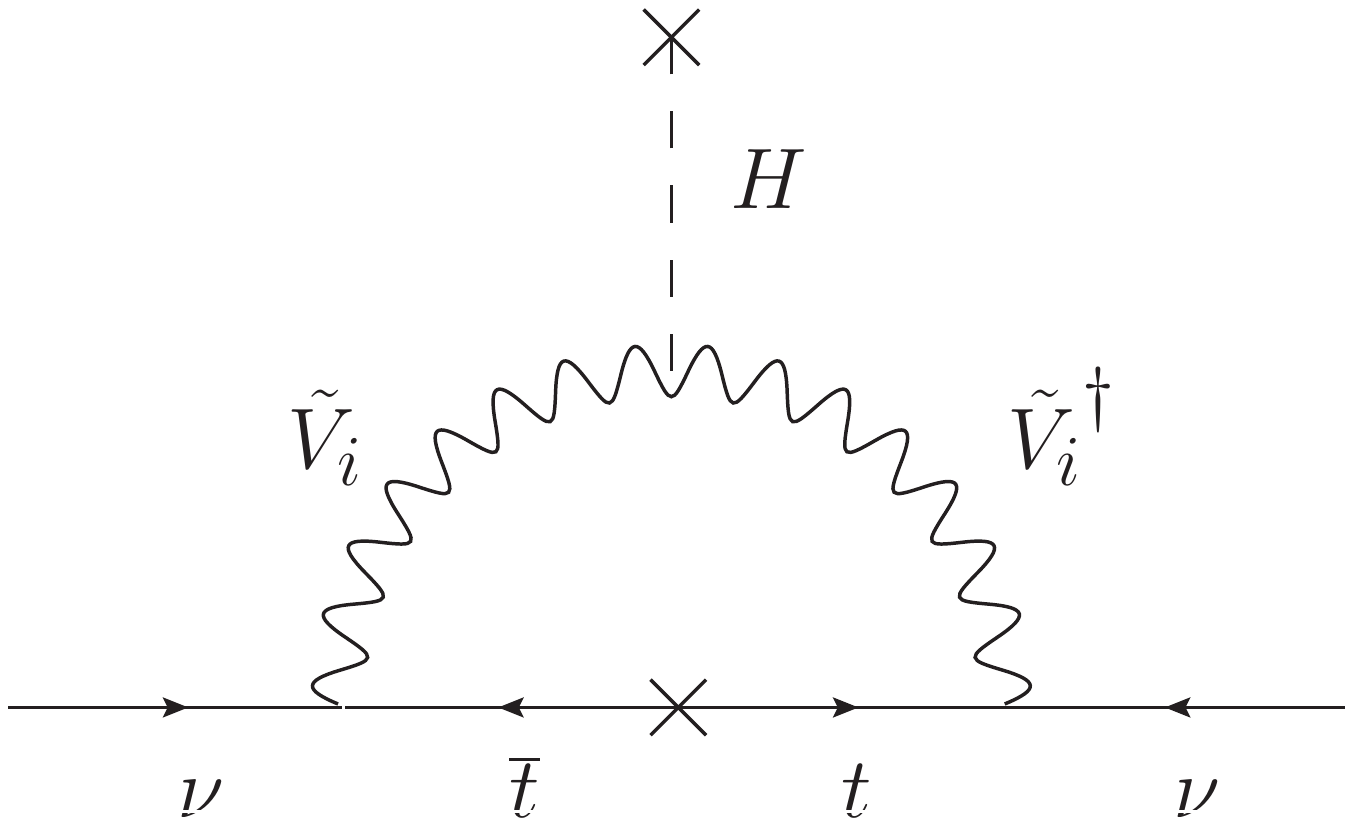}
\vspace{-8cm} 
\caption{One-loop Majorana neutrino mass generated by Higgs-leptoquark mixing.}
\label{nudiag}
\end{figure}

Since two leptoquarks with couplings to up-type quarks and neutrinos are present, their $\Delta L =2$ mixing induced by the Higgs boson interaction $h_{V} H i \tau_2 V_{1/2}^{\mu} V_{0\mu}^{\dagger}$ generates Majorana neutrino masses at the one-loop level as depicted in Fig. \ref{nudiag}. 

The magnitude of the neutrino mass depends on the leptoquark mixing, governed by the dimensionful parameter $h_V$,  and on the leptoquark couplings $\lambda^{L,R}$. Explicitly, the contribution to the Majorana neutrino mass from $V_{0,\mu}$ and $V_{1/2,\mu}$ is given by \cite{AristizabalSierra:2007nf}
\begin{align}
  M^{\nu}_{i i^{\prime}} &= \frac{3}{16 \pi^2} \sum_{j=1,2} \sum_{k=u,c,t} m_k B_0(0,m_k^2,m_{V_j}^2) R_{j1} R_{j2} \nonumber \\
&\times \left[ \lambda^R_{k i} \lambda^L_{k i^\prime} + \lambda^R_{k i^\prime} \lambda^L_{k i} \right]\,, \label{numass}
\end{align}
where $m_{V_j}$ is the mass of the leptoquark $V_j$, $m_k$ is the quark mass and $R_{jl}$ is the mixing matrix diagonalizing the leptoquark mass matrix, while $B_0$ denotes the finite part of the Passarino-Veltman function
\begin{align}
B_0(0,m_k^2,m_{V_j}^2) = \frac{m_k^2 \log(m_k^2) - m_{V_j}^2 \log (m_{V_j}^2)}{m_k^2 - m_{V_j}^2}\,. \label{PV}
\end{align}
A few comments regarding the loop regularization are in order. As seen, for instance, in the unitary gauge
\begin{align}
  \frac{-i}{k^2 - M_{V}^2} \left[g_{\mu\nu} + \frac{k_{\mu}k_{\nu}}{M_V^2} \right]\,,
 \end{align}
the vector leptoquark propagator causes divergences that result in a bad UV behavior. Analogous to the Higgs and the $W^{\pm}$ bosons in the SM, a heavy Higgs giving masses to the leptoquarks can cancel these divergences. The details, however, depend on the specific UV completion. An example where neutrino masses are mediated by a massive gauge boson is given in Ref. \cite{Boucenna:2014dia}. Here, massive bosons emerge through the breaking of a $SU(3)_C \otimes SU(3)_L \otimes U(1)_X$ gauge group and the ultraviolet behavior is well defined.
Another example is shown in Ref. \cite{Ma:2012xj} for an $SU(2)_N$ extension of the SM, where the $SU(2)_N$ gauge bosons generate a nonzero neutrino mass.

The remaining infinities contained in the Passarino-Veltman function drop out when summing over both leptoquarks considered in our analysis. The function $B_0$ in Eq. (\ref{PV}), therefore, takes into account only the finite part of the Passarino-Veltman integral. 

Stringent constraints can arise if the UV completion does not entail additional particles to cancel the divergences of the vector-boson propagator. Such limits, e.g., from radiative charged lepton decays $l \rightarrow l^{\prime} \gamma$, are discussed in Ref. \cite{Barbieri:2015yvd} based on the vector leptoquark $(3,1)_{2/3}$ in a $U(2)^5$ flavor model.

Using the leptoquark patterns discussed in Eqs. (\ref{FNpatt1})--(\ref{FNpatt3}), we can estimate the absolute neutrino mass scale generated by the leptoquark couplings. Since the patterns are strongly hierarchical in terms of quark families, we need only consider the dominating top quark contribution to $M^{\nu}_{i i^{\prime}}$. Hence, we obtain
\begin{align}
M_{i i^{\prime}}^{\nu} &\approx \underbrace{\frac{3}{32 \pi^2} m_t \sin 2 \alpha \Delta B_0}_{\equiv a} \left[ \lambda^R_{t i} \lambda^L_{t i^\prime} + \lambda^R_{t i^\prime} \lambda^L_{t i} \right]\,, \nonumber \\
\Delta B_0 &\equiv B_0(0,m_t^2,m_{V_{1/2}}^2) - B_0(0,m_t^2,m_{V_{0}}^2),\,
\end{align}
and the neutrino mass eigenstates 
\begin{align}
 m^{\nu}_1 &= 0, \\
 \frac{m^{\nu}_{2(3)}}{a} &= \sum_{i} \lambda^L_{ti} \lambda^R_{ti} \substack{- \\ (+)} \sqrt{\sum_{i} \left(\lambda^L_{ti}\right)^2 \sum_{i} \left(\lambda^R_{ti}\right)^2} \label{num}
\end{align}
with $i=e,\mu,\tau$. Note that one eigenvalue is exactly zero if either only down-type or up-type quarks generate neutrino masses. Hence, the model predicts a normal neutrino mass hierarchy with a small effective Majorana mass relevant for $0\nu\beta\beta$.

\begin{figure}[tbh]
\centering
\includegraphics[width=0.47\textwidth]{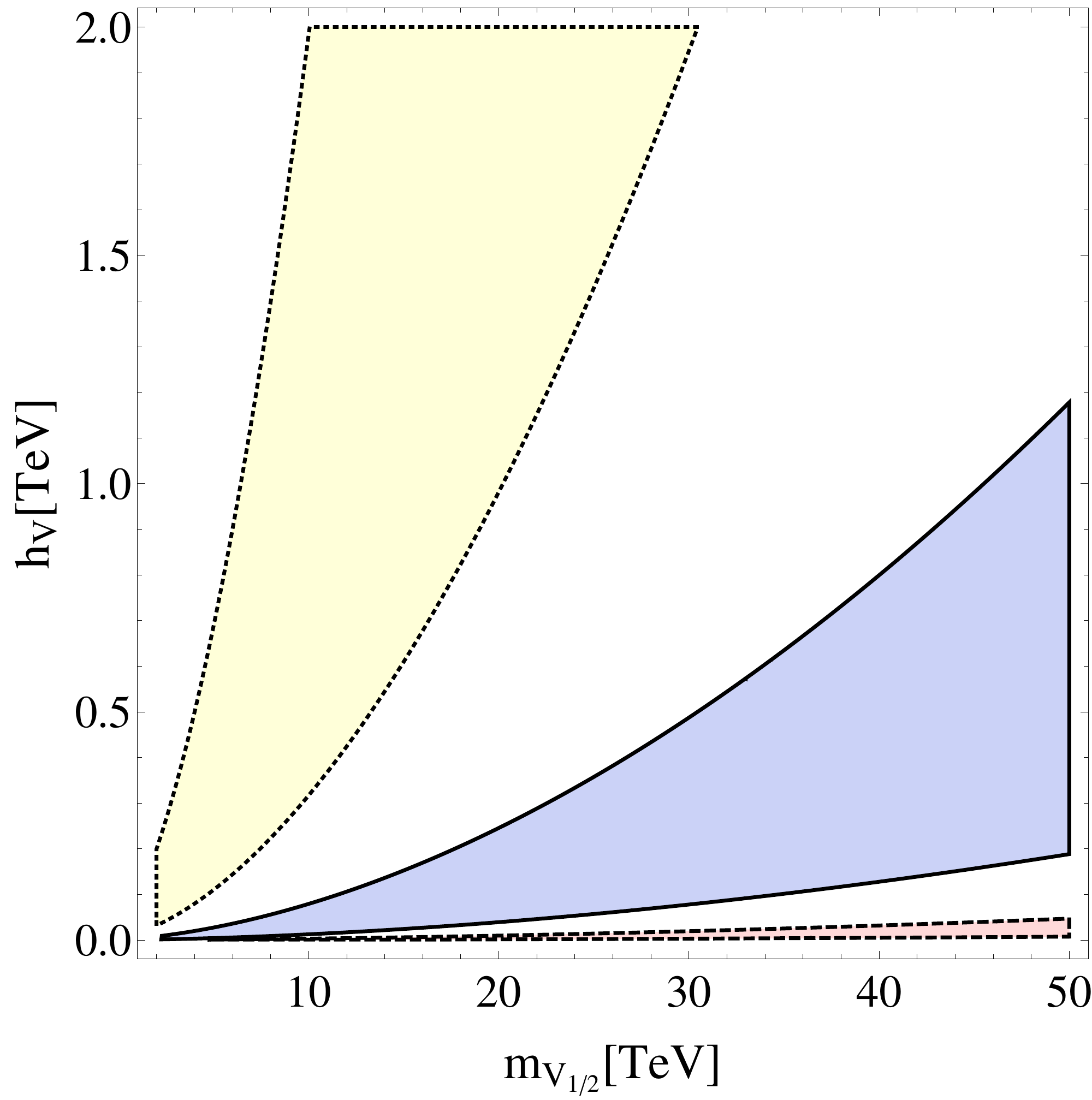}
\caption{Allowed regions of the trilinear leptoquark-Higgs coupling $h_V$ and $m_{V_{1/2}}$ requiring that the largest neutrino mass eigenstate $m_{\nu} \lesssim 0.3$ eV and $m_{V_0} \simeq 1$ TeV. The three distinct regions correspond to different powers of the dominating coupling $\lambda^R_{t\nu} \simeq \epsilon^8$ (red, dashed), $\epsilon^{10}$ (blue, solid) and $\epsilon^{12}$ (yellow, dotted).}
\label{mnu}
\end{figure}

\begin{figure}[tbh]
\centering
\includegraphics[width=0.47\textwidth]{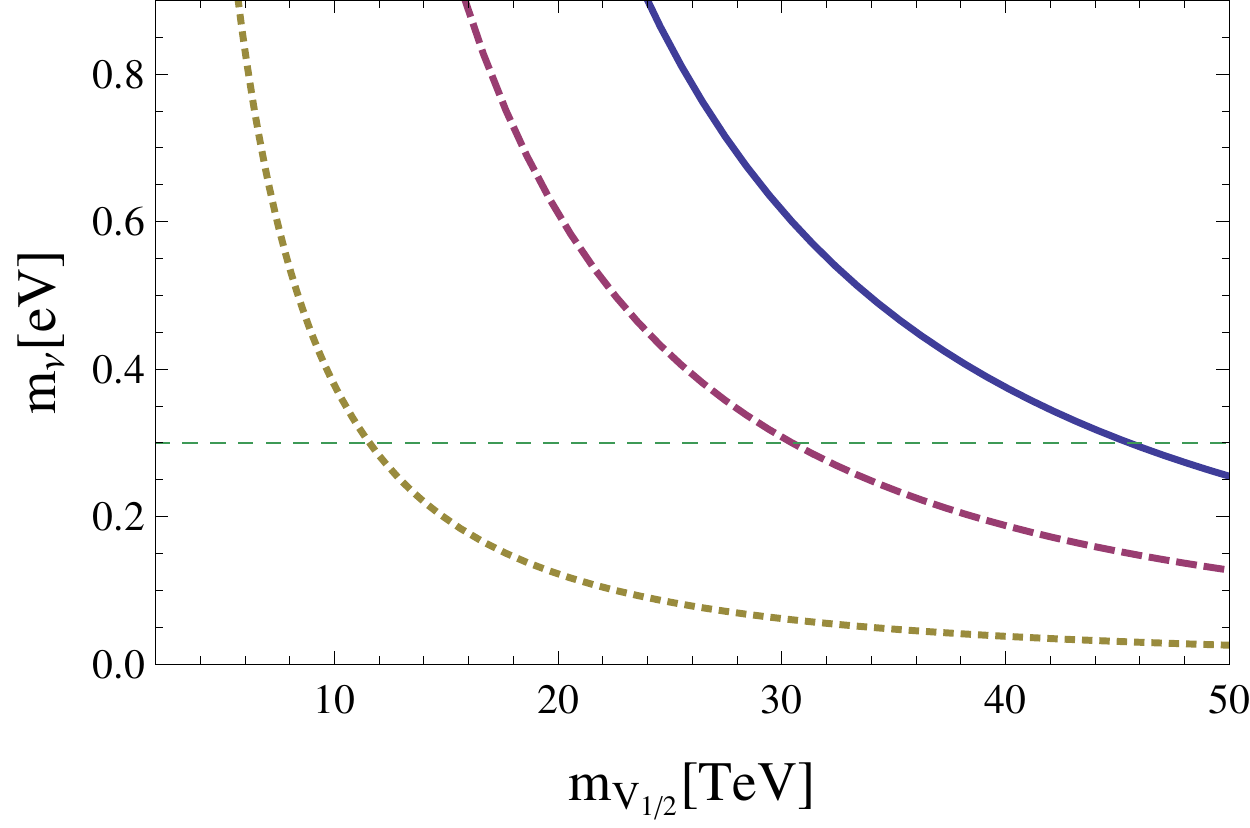}
\caption{Heaviest neutrino mass eigenstate as a function of $m_{V_{1/2}}$ for $m_{V_0}=1$ TeV, $\Lambda^R_{t\nu} \simeq \epsilon^{10}$ and $h_V = 1$ TeV (blue, solid), $0.5$ TeV (red, dashed), and $0.1$ TeV (yellow, dotted). The horizontal, dashed line defines a conservative upper limit on the heavy neutrino mass.}
\label{mnu2}
\end{figure}

Inserting Eqs. (\ref{FNpatt1})--(\ref{FNpatt3}) yields
\begin{align} 
M_{i i^{\prime}}^{\nu} \propto a \cdot \left( 
\begin{array}{ccc}
\epsilon^{16} & \epsilon^{14} & \epsilon^{13} \\ 
\epsilon^{14} & \epsilon^{12} & \epsilon^{11} \\ 
\epsilon^{13} & \epsilon^{11} & \epsilon^{10}
\end{array}
\right),\quad 
\text{and} \quad m^{\nu}_3 \sim a \cdot \epsilon^{10}\,.\label{estimate}
\end{align} 
Therefore, the factor $a$ must be sufficiently small to push the neutrino mass scale below eV, which is achieved by virtue of small leptoquark mixing. In the limit of small $\alpha$ the parameter $a$ can be approximated as
\begin{align}
 a \approx \frac{3}{16 \pi^2} m_t \frac{h_V v_{\text{SM}}}{\Delta m_V^2} \log \left[\frac{m_{V_{1/2}}^2}{m_{V_0}^2}\right]\,,
\end{align}
implying 
\begin{align} \label{lesseq}
 \frac{h_V v_{\text{SM}}}{\Delta m_V^2} \log \left[\frac{m_{V_{1/2}}^2}{m_{V_0}^2}\right] \lesssim 0.9 \times 10^{-3} 
\end{align}
to make neutrino masses sufficiently light. The smallness of $a$ can be attributed to the smallness of the dimensionful  coupling $h_V$ or a large mass splitting $\Delta m_V^2$ of the contributing leptoquarks. Possible solutions of Eq. (\ref{lesseq}) are depicted in Fig. \ref{mnu} for different powers of $\lambda^R \sim \epsilon^8, \epsilon^{10},$ and $\epsilon^{12}$. In Fig. \ref{mnu2} we plot $m^{\nu}_3$ in terms of $m_{V_{1/2}}$ for $\lambda^R \sim \epsilon^{10}$ and $h_V = 0.1, 0.5,$ and $1\,$TeV, showing that light neutrino masses favor a large leptoquark mass splitting with natural values of $h_V$.  

Since one neutrino mass eigenstate is exactly zero, one can solve the eigenvalue equation $M_{\nu} v_0 = 0$ with 
\begin{align}
v_0^T = \frac{(1,-w,w^{\prime})}{\sqrt{1+w^2+w^{\prime 2}}}
\end{align}
to obtain analytical expressions for the neutrino mixing angles as a function of the leptoquark couplings $\lambda^{L,R}_{t\nu}$, assuming the charged leptons are approximately diagonal \cite{AristizabalSierra:2007nf}. It is 
\begin{align}
w &= \frac{\lambda^R_{t\tau}\lambda^L_{te}-\lambda^R_{te}\lambda^L_{t\tau}}{\lambda^R_{t\tau}\lambda^L_{t\mu}-\lambda^R_{t\mu}\lambda^L_{t\tau}} \approx t_{12} \frac{c_{23}}{c_{13}} + t_{13}s_{23}\,,\\
w^{\prime} &= \frac{\lambda^R_{t\mu}\lambda^L_{te}-\lambda^R_{te}\lambda^L_{t\mu}}{\lambda^R_{t\tau}\lambda^L_{t\mu}-\lambda^R_{t\mu}\lambda^L_{t\tau}} \approx t_{12} \frac{s_{23}}{c_{13}} - t_{13}c_{23}\,,
\end{align}
where, $s_{23}=\sin \theta_{23}$ etc. with the PMNS angles $\theta_{12}$,$\theta_{23}$,$\theta_{13}$. Hence, to explain large PMNS mixing $w$ and $w^{\prime}$ both should be nonzero and sizable. By evaluating $w$ and $w^{\prime}$ for Eqs. (\ref{FNpatt1})--(\ref{FNpatt3}) we find that their values depend heavily on the undetermined $\mathcal{O}(1)$ FN parameters
\begin{align}
w=\frac{y_{31}-y_{13}}{y_{32}-y_{23}}\epsilon^2\,,\qquad w^{\prime}=\frac{y_{21}-y_{12}}{y_{32}-y_{23}}\epsilon^3\,,
\end{align}
where $y_{ij}$ denote products of $\mathcal{O}(1)$ coefficients from $\lambda^{L,R}$. Because of possible cancellation in the denominator, $w$ and $w^{\prime}$ can oscillate quickly with small changes of the $\mathcal{O}(1)$ parameters, explaining also large neutrino mixing easily by permitting some extent of tuning.

With many free $\mathcal{O}(1)$ FN parameters to match to only five physical observables ($\Delta m_{\text{atm}}^2, \Delta m_{\text{sol}}^2, \theta_{12}, \theta_{13}, \theta_{23})$, the system is underconstrained and has many viable solutions. On condition that all coefficients in Eq. (\ref{ml}) are approximately $\mathcal{O}(1)$, the benchmark point
\begin{align} \begin{array}{lll}
 &\lambda_{te}^L \approx 5.1\,\epsilon^3\,, \qquad & \lambda_{te}^R \approx 3.0\,\epsilon^{13}\,, \\
 &\lambda_{t\mu}^L \approx 1.4\,\epsilon\,,\qquad & \lambda_{t\mu}^R \approx 2.1\,\epsilon^{11}\,, \\
 &\lambda_{t\tau}^L \approx 0.2\,, \qquad & \lambda_{t\tau}^R \approx -0.8\,\epsilon^{10}\,, 
 \end{array}
\end{align}
provides an excellent fit to neutrino oscillation data, yielding
\begin{align}
 &\Delta m_{\text{atm}}^2 = 2.5 \times 10^{-3}\,\text{eV}^2, \quad \Delta m_{\text{sol}}^2 = 7.6 \times 10^{-5}\,\text{eV}^2, \nonumber \\
 &\theta_{12} = 33.3^{\circ},\qquad \theta_{13} = 8.5^{\circ},\qquad \theta_{23} = 42.0^{\circ}\,. 
\end{align}

Further limits on $\Delta L=2$ lepton-number-violating leptoquark couplings also arise from $0\nu\beta\beta$ experiments, which can be even more stringent than LHC searches \cite{Hirsch:1996ye,Helo:2013ika}. 
The mixing of $V_0$ and $V_{1/2}$ induced by the SM Higgs boson generates the operator \cite{Hirsch:1996ye}
\begin{align} 
	\lambda^L_{de} \lambda^R_{u\nu}
	\frac{h_V v_\text{SM}}{m_{V_{1/2}}^2 m_{V_{0}}^2}
	\left[\overline{\nu} P_R e^c \right] \left[\overline{u} P_R d \right]\,.
\end{align}
Given the strong suppression of first generation couplings combined with the extra suppression of $\lambda^R$, the $0\nu\beta\beta$ bound is negligible in this framework. By contrast, scalar leptoquarks with inverse hierarchical patterns can reduce the $0\nu\beta\beta$ half-life considerably, allowing for an observation of the $0\nu\beta\beta$ decay in the near future \cite{Pas:2015hca}.

\section{750 GeV diphoton excess}
\label{secdiph}
Recently, the ATLAS and CMS collaborations reported an excess in the diphoton spectrum near $750$ GeV with $3.9\,\sigma$ and $2.6\,\sigma$ local significance \cite{ATLAS:2015abc,CMS:2015dxe}. The signal hints at a potential resonance with spin 0 or 2 and strongly enhanced branching ratios into gluons and photons.

A plethora of explanations has been considered by various authors since the announcement of the excess, among them also leptoquark mediators. While pure scalar leptoquark solutions face difficulties regarding unitarity, vector leptoquarks can explain the signal rather elegantly thanks to a sizable loop factor. The beauty of the vector leptoquark solution is that it does not come with numerous exotic fermions to artificially enhance the diphoton decay mode.

The vector leptoquarks in our model can interact with the scalar resonance $\chi$ through the hypothetical interaction
\begin{align}
\mathcal{L}_{V \chi} = \kappa_{V_i}\chi V_{\mu,i}^{\dagger} V^{\mu}_i\, + \text{H.c.}\,,
\end{align}
 where $i=0,\frac{1}{2}$. $\kappa_{V_i}$ is a dimensionful parameter whose scale thus far is undetermined, however bounded from above by unitarity constraints. The scale where the theory breaks down can be roughly inferred from elastic $V_{i,\mu} V^{i,\mu} \rightarrow V_{i,\mu} V^{i,\mu}$ scattering, given by $\sqrt{s} \sim 4\sqrt{\pi} m^2_{V_i}/|\kappa_{V_i}|$ \cite{Murphy:2015kag}. In the following we will assume natural TeV-scale values for $\kappa_{V_i}$ to comply with perturbative unitarity.

\begin{figure}[tbh]
\centering 
\includegraphics[width=0.6\textwidth]{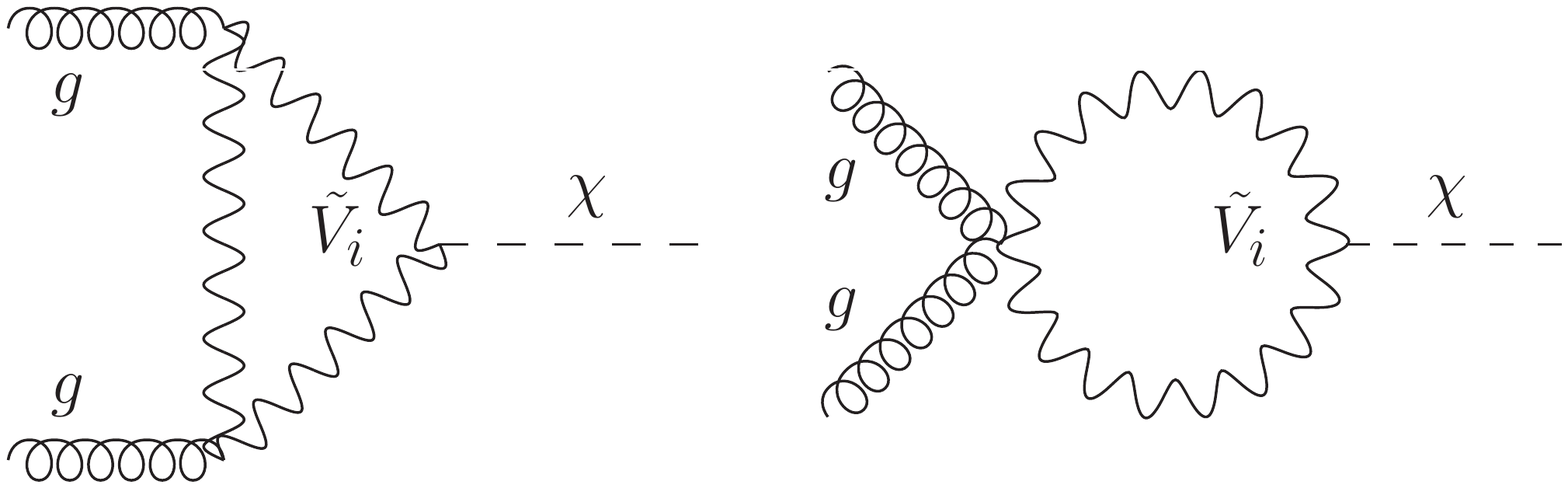}
\vspace{-4.8cm}
\caption{Dominating diagrams contributing to $\sigma(pp\rightarrow\chi)$.}
\label{chigg}
\end{figure}

\begin{figure}[tbh]
\centering
\includegraphics[width=0.6\textwidth]{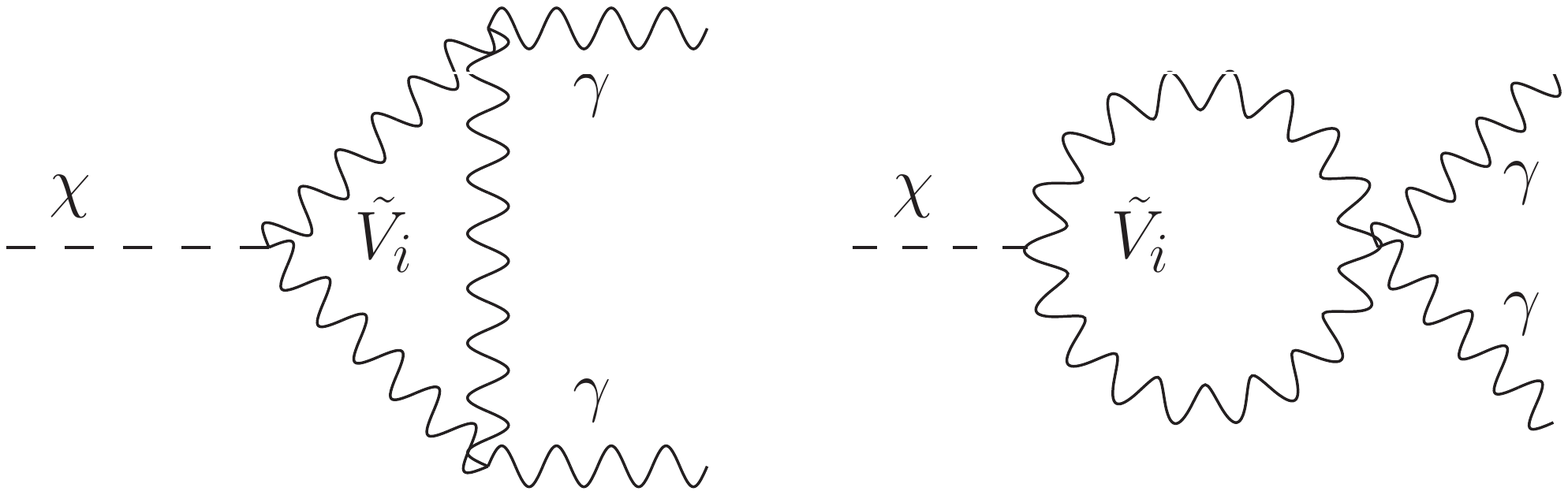}
\vspace{-4.8cm}
\caption{Diagrams contributing to $\Gamma(\chi \rightarrow \gamma\gamma)$.}
\label{chigg2}
\end{figure}

The total cross section $\sigma$ is a product of the $\chi$ production and its subsequent decay rate into two photons. $\chi$ production from $q\overline{q}$ initial states is possible, however, strongly suppressed either by small leptoquark couplings (cf. Eqs. (\ref{FNpatt1})--(\ref{FNpatt3})) or small values of the parton distribution functions at $\sqrt{s}=13$ TeV \cite{Franceschini:2015kwy}. The only partly competitive channel in terms of luminosity, $d\overline{d} \rightarrow \chi$, is additionally suppressed compared to gluon fusion by a factor of $|\lambda^L_{d\tau}|^2 = \epsilon^6$ due to the FN symmetry. Hence, assuming that $\chi$ is predominantly produced via gluon fusion we obtain \cite{Leike:1998wr}
\begin{align}
\sigma(pp \rightarrow \chi \rightarrow \gamma\gamma) &= \frac{\pi^2}{8s} \frac{\Gamma (\chi \rightarrow \gamma\gamma)}{\Gamma_{\text{tot}}} \nonumber \\ 
&\times \frac{\Gamma(\chi \rightarrow gg)}{m_{\chi}}f_{gg} (m_{\chi}^2/s) 
\end{align}
with 
\begin{align}
 \Gamma(\chi \rightarrow gg) &= \frac{\alpha_s^2 m_{\chi}^3 K^{gg}}{128 \pi^3} \left| \sum_i \frac{\kappa_{V_i} A_1(\tau_{V_i})}{m_{V_i}^2} \right|^2, \\
 \Gamma(\chi \rightarrow \gamma\gamma) &= \frac{\alpha_e^2 m_{\chi}^3}{256 \pi^3} \left| \sum_i \frac{\kappa_{V_i} N_c Q_{V_i}^2 A_1(\tau_{V_i})}{m_{V_i}^2} \right|^2\,,
\end{align}
where $i=0,\frac{1}{2}$. Henceforth, we will denote $\sigma(pp \rightarrow \chi \rightarrow \gamma\gamma)$ by $\sigma_{\gamma\gamma}$, $\Gamma(\chi \rightarrow gg)$ by $\Gamma_{gg}$ and $\Gamma(\chi \rightarrow \gamma\gamma)$ by $\Gamma_{\gamma\gamma}$. We furthermore approximate $\Gamma_{tot} \approx \Gamma_{gg}$. $K^{gg} \approx 1.5$ accounts for higher-order QCD corrections, $N_c=3$ for the vector leptoquarks running in the loop and $\alpha_s$ is the strong coupling constant. $A_1(\tau)$ denotes a loop factor for a spin-1 particle given by \cite{Djouadi:2005gj}
\begin{align}
 A_1(\tau) &= \frac{1}{\tau^2}\left[2\tau^2 + 3\tau +3(2\tau-1)\arcsin^2 \sqrt{\tau}\right]\,,
\end{align}
and $\tau_{V_i}=m_\chi^2/(4m_{V_i}^2) < 1$. The loop factor $A_1(\tau)$ was originally computed in Ref. \cite{Gunion:1989we} to account for the $W^{\pm}$ gauge boson contribution to the radiative decay $h \rightarrow \gamma \gamma$ in the SM. The unphysical degrees of freedom of the vector bosons can cause loop divergences that have to be dealt with. These divergences were regularized using the nonlinear $R_{\xi}$ gauge \cite{Gavela:1981ri}, in which the vector-boson propagator reads
 \begin{align}
  \frac{-i}{k^2 - M_{V}^2} \left[g_{\mu\nu} + \frac{(\xi-1)k_{\mu}k_{\nu}}{k^2-\xi M_V^2} \right]\,.
 \end{align}
It is shown in Ref. \cite{Gavela:1981ri} that all divergences cancel out separately in the vector-boson and the Faddeev-Popov ghost sector, resulting in a finite and gauge-independent theory.

$A_1(\tau)$ acquires large values for vector leptoquarks compared to scalar particles. For comparison, the spin-0 and spin-1/2 loop factors read
\begin{align}
 A_0(\tau) &= -\frac{1}{\tau^2}\left[\tau - \arcsin^2 \sqrt{\tau}\right]\,, \\
 A_{1/2}(\tau) &= \frac{2}{\tau^2}\left[ \tau + (\tau-1)\arcsin^2 \sqrt{\tau} \right]\,,
\end{align}
respectively. Assuming masses ranging from $\sim 0.8$ to $50$ TeV, the loop factors remain near constant and
\begin{align}
 \frac{|A_1(\tau)|}{|A_0(\tau)|} \approx 20\,, \qquad  \frac{|A_1(\tau)|}{|A_{1/2}(\tau)|} \approx 5,
\end{align}
in the relevant mass region.

The gluon luminosity function $f_{gg}$, evaluated at $\sqrt{s}=13$ TeV using MSTW2008 \cite{Martin:2009iq} leads to
\begin{align}
f_{gg}= \int_{m_{\chi}^2/s}^{1} f_g(x) f_g (m_{\chi}^2/ (x s) ) \frac{\text{d}x}{x}  = 2141.7\,, 
\end{align}
where $f_g$ is the gluon distribution function. Depending on the dimensionful couplings $\kappa_{V_i}$, typical values of $\Gamma_{gg}/m_\chi$ and $\Gamma_{\gamma\gamma}/m_\chi$ are $\mathcal{O}(10^{-4})$ and $\mathcal{O}(10^{-6})$, respectively. In our setup, at the benchmark point $\kappa_{V_i} = \frac{4}{3} m_{V_i}$, $m_{V_0} = 1$ TeV  and $m_{V_{1/2}}=20$ TeV we have
\begin{align}
 \frac{\Gamma_{gg}}{m_\chi} \simeq 2 \cdot 10^{-4}\,,\quad  \frac{\Gamma_{\gamma\gamma}}{m_\chi} \simeq 8 \cdot 10^{-7}\,,\quad \sigma_{\gamma\gamma}\simeq 4~\text{fb}\,.
\end{align}
Therefore, the estimated dijet cross section at 13~TeV is 4~pb, leading to a cross section $\simeq 0.8$~pb at 8~TeV. Currently the ATLAS and CMS collaborations do not provide dijet limits at $\sqrt{s} = 13$~TeV for resonance masses below 1~TeV. The $\sqrt{s} = 8$~TeV ATLAS and CMS analyses presented in Refs. \cite{Khachatryan:2015sja, Aad:2014aqa} set a limit of $\sigma_{jj} < 1$~pb for a 1~TeV resonance coupling dominantly to $gg$. For a mass of 750~GeV the limit shown by ATLAS is of the order of 10~pb. Hence within the interesting region of parameter space considered here, the dijet limits are satisfied.

\begin{figure}[t]
\includegraphics[width=0.46\textwidth]{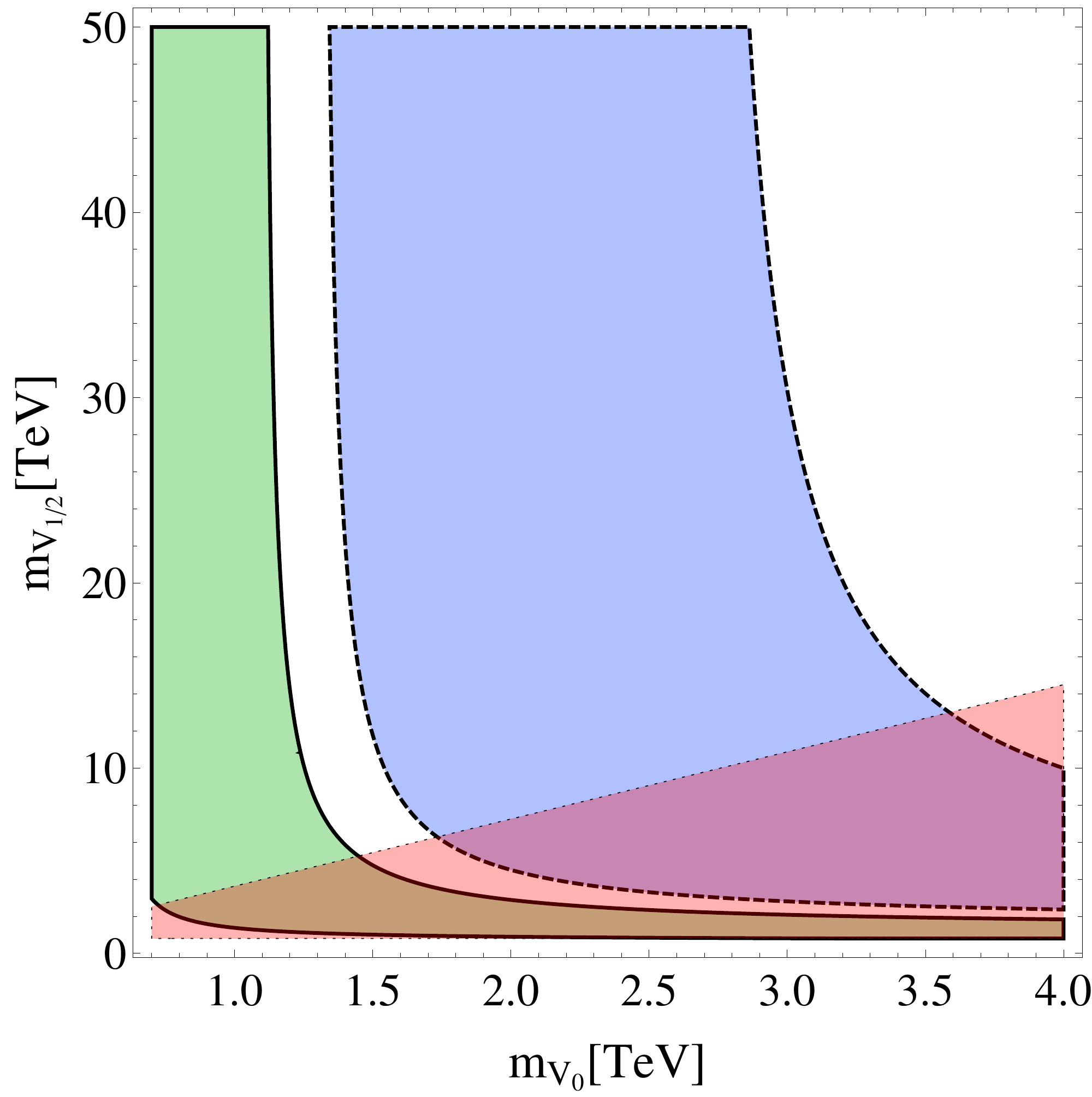}
\caption{Parameter regions yielding $\sigma_{\gamma\gamma} \in (3,13)$ fb as measured by ATLAS and CMS, where $\sigma_{\gamma\gamma}$ is shown as a function of the leptoquark masses $m_{V_0}$ and $m_{V_{1/2}}$ for different values of the dimensionful couplings $\kappa_{V_i} = \frac{10}{3} m_{V_i}$ (blue, dashed) and $\frac{4}{3} m_{V_i}$ (green, solid). The constraint (\ref{eq:DibosonConstraint}) with $r=0.28$ from $WW$, $ZZ$ and $Z\gamma$ limits is superimposed in red.}
\label{regionsa}
\end{figure}
As $V_0$ and $V_{1/2}$ carry hypercharge, they necessarily decay via $\chi \to Z \gamma$ and $\chi \to ZZ$. Limits on these final states from experimental collaborations already exist. Here we take a rather simplistic viewpoint and assess the viability of our scenario without explicitly calculating the cross sections for $Z \gamma, ZZ$ final states. This can be done by estimating the ratios of $\chi$ partial widths. The partial widths $\chi \to Z \gamma$ and $\chi \to ZZ$ are suppressed by $2\tan^2 \theta_W$ and $\tan^4\theta_W$, respectively, compared to $\Gamma_{\gamma\gamma}$ and existing bounds on these channels can be easily evaded. More importantly, $V_{1/2}$ is an $SU(2)$ doublet with enhanced rates $\Gamma_{Z\gamma}/\Gamma_{\gamma\gamma} \approx 2/\tan^2 \theta_W$, $\Gamma_{ZZ}/\Gamma_{\gamma\gamma} \approx 1/\tan^4 \theta_W$. In addition, the decay to two $W$ bosons will be possible as well with a strongly enhanced rate $\Gamma_{WW}/\Gamma_{\gamma\gamma} \approx 2/\sin^2 \theta_W$ \cite{Franceschini:2015kwy}. The experimental limits are satisfied if
\begin{align} \label{eq:DibosonConstraint}
 \frac{|\kappa_{V_{1/2}}|}{m^2_{V_{1/2}}} < r \times \frac{|\kappa_{V_0}|}{m^2_{V_0}},
\end{align}
with $r \approx 3.1$ if the $\kappa$ couplings have the same sign and $r \approx 0.28$ if they have opposite signs. The difference arises due to constructive or destructive interference from the contribution of $SU(2)$ and $U(1)$ coupling components to the decay widths. We use this constraint in order to quantify the impact of diboson final-state limits in our analysis.

The width of $\chi$ is dominated by the decay to gluons and it is typically small, $\Gamma_\text{tot} \approx \Gamma_{gg} \approx 0.3$ GeV. We make no attempt to explain a potentially large width as suggested by ATLAS within this setup.

In the following we determine the allowed parameter ranges of $\kappa_{V_i}$ and $m_{V_i}$ to reproduce the total cross sections measured by ATLAS and CMS in the diphoton channel near 750 GeV
\begin{align} \label{sigmadiph}
 \sigma_{\text{ATLAS}} = (10 \pm 3)~\text{fb}\,,\qquad \sigma_{\text{CMS}} = (6 \pm 3)~\text{fb}\,.
\end{align}

\begin{figure}[t]
\includegraphics[width=0.47\textwidth]{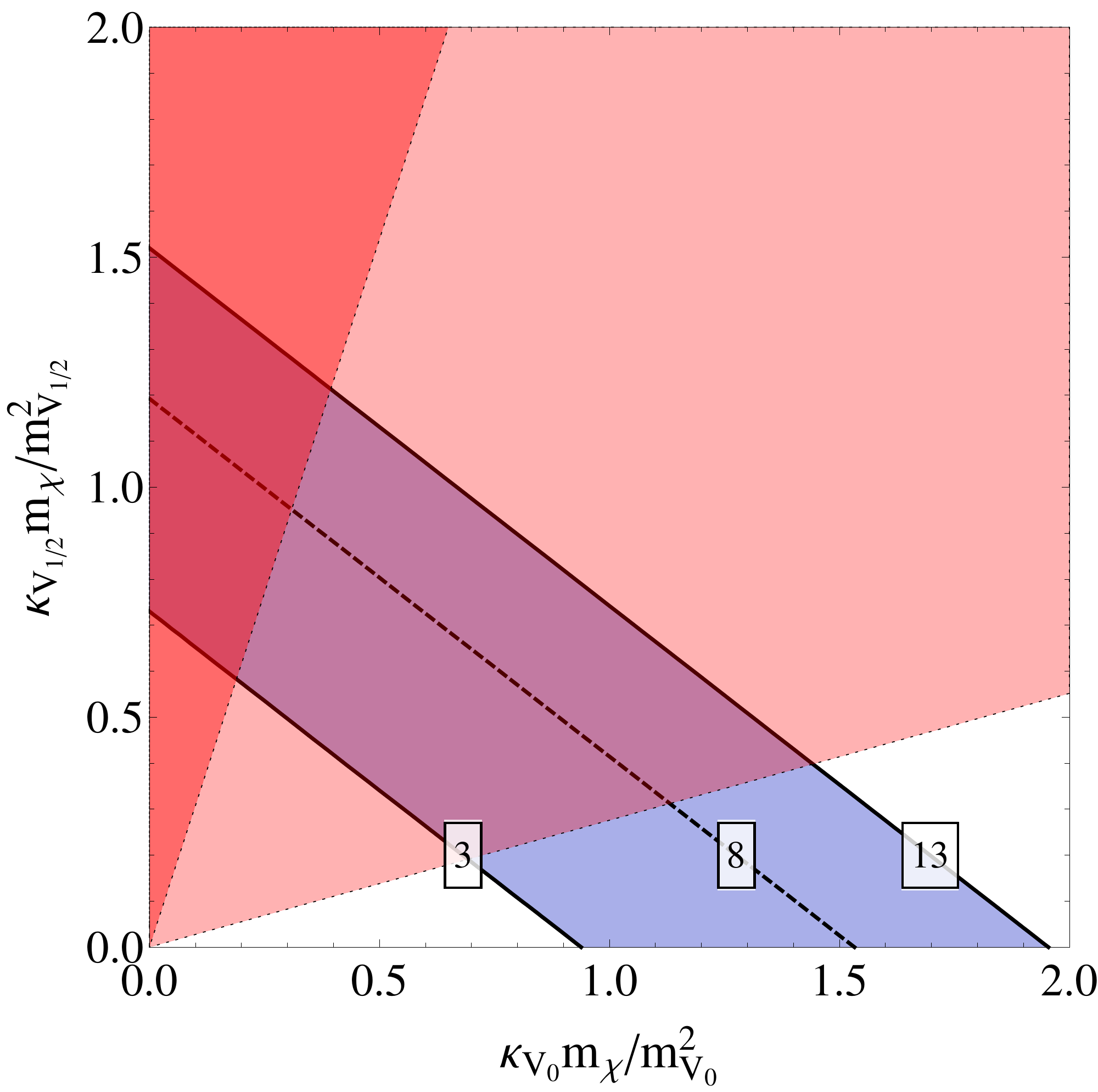}
\caption{Parameter regions yielding $\sigma_{\gamma\gamma} \in (3,13)$~fb as measured by ATLAS and CMS. $\sigma_{\gamma\gamma}$ as a function of the effective parameters $\kappa_{V_i} m_{\chi} / m_{V_i}^2$. The lines denote values of constant $\sigma_{\gamma\gamma}$ in fb. The constraint Eq.~\eqref{eq:DibosonConstraint} with $r=0.28$ (light shade) and $r=3.1$ (dark shade) from $WW$, $ZZ$ and $Z\gamma$ limits is superimposed in red.}
\label{regionsb}
\end{figure}
Taking into account that $m_{V_{0}} \sim 1$~TeV is needed to reproduce the $R_K$ and $R_D$ data, we obtain the allowed parameter regions displayed in Figs.~\ref{regionsa} and \ref{regionsb} as a function of the dimensionful couplings $\kappa_{V_i}$ and the leptoquark masses $m_{V_i}$, respectively. The parameter space favoring the diphoton cross section opens up notably if the second leptoquark is much heavier, yielding a large $\Delta m_{V}^2$ that is also favored by neutrino mass generation. Figure~\ref{regionsa} represents the $\sigma_{\gamma\gamma}$ in the range of $3 - 13$ fb, for two different values of dimensionful couplings $\kappa_{V_i}$. The parameter space excluded by $WW, ZZ, Z\gamma$ searches is depicted in red. This constraint is derived using Eq.~\eqref{eq:DibosonConstraint}. In Fig.~\ref{regionsa} only the more stringent constraint applicable if the $\kappa$ couplings have opposite sign is shown; the case of same sign exhibits no appreciable constraint.

\begin{figure}[t]
\includegraphics[width=0.47\textwidth]{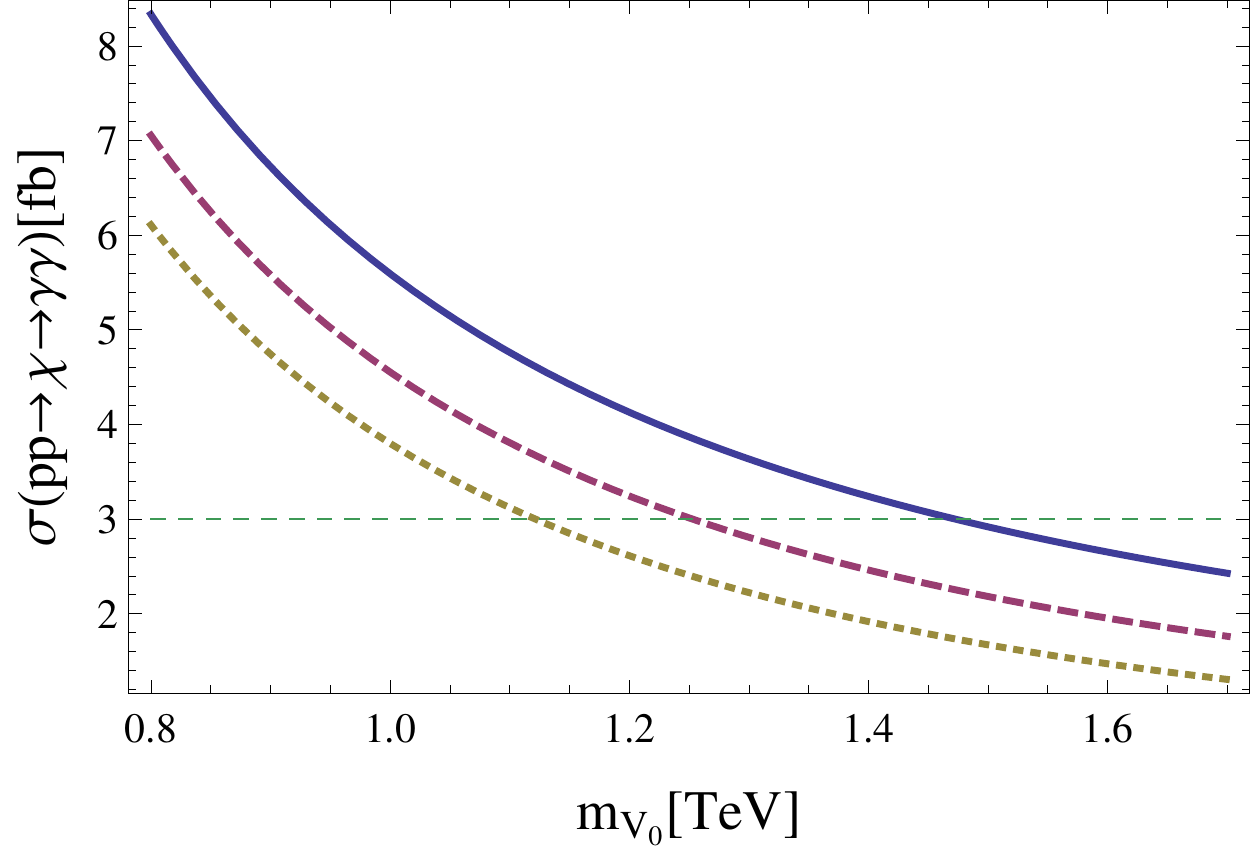}
\caption{$\sigma(pp \to \chi \to \gamma\gamma)$ as a function of $m_{V_0}$ for $m_{V_{1/2}} = 5$ TeV (blue, solid), $10$ TeV (red, dashed), $50$ TeV (yellow, dotted) with $\kappa_{V_i} = \frac{4}{3}m_{V_i}$. The horizontal, dashed lines correspond to the lower limit given by the ATLAS and CMS diphoton measurements.}
\label{lines}
\end{figure}

\begin{figure}[t]
\includegraphics[width=0.47\textwidth]{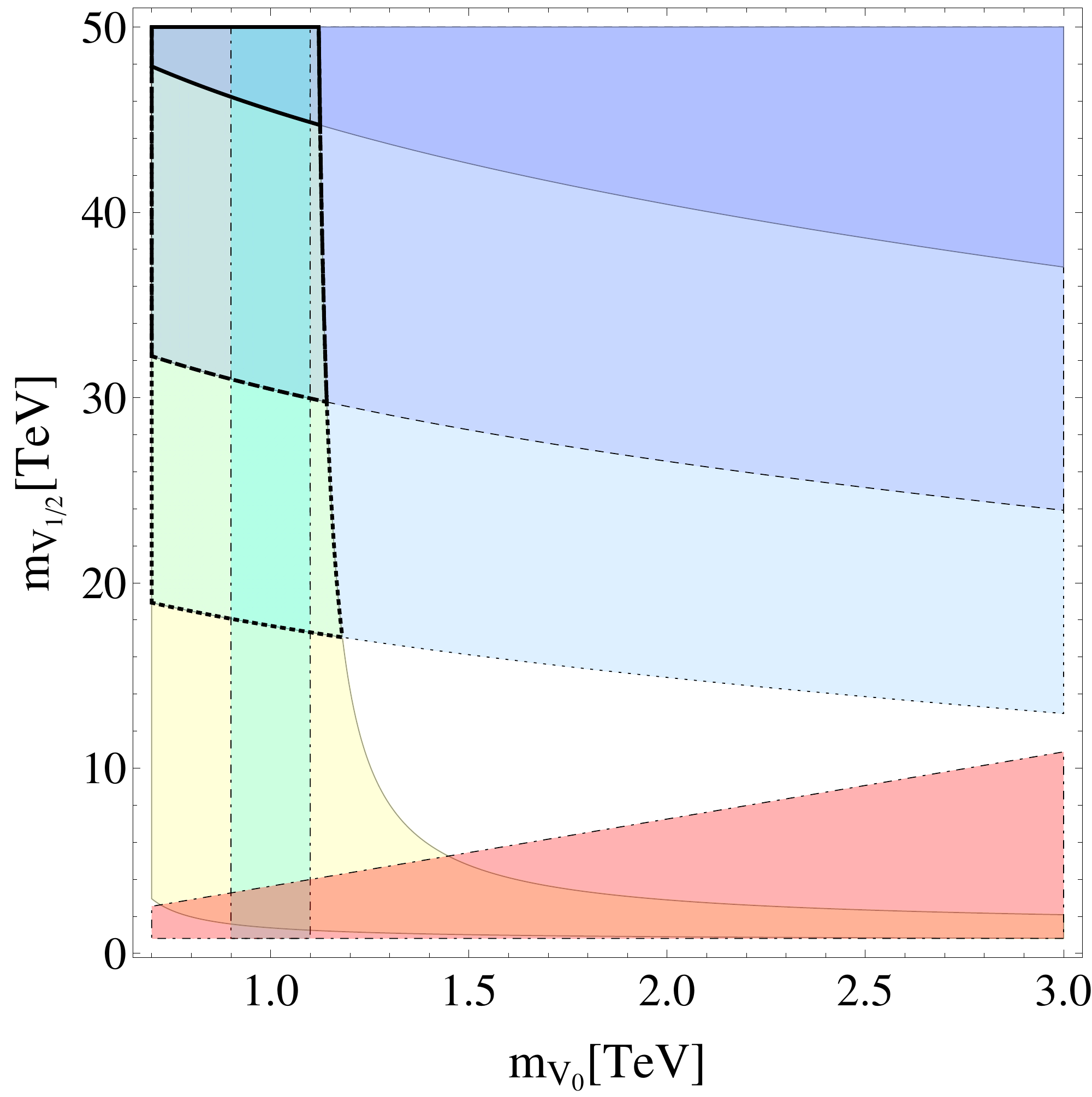}
\caption{Fit results of the 750 GeV diphoton excess (yellow) superimposed with constraints from neutrino mass generation (blue) in the leptoquark parameter space for $\kappa_{V_i} = \frac{4}{3}m_{V_i}$ and $h_V = 0.2$ TeV (dotted), $0.5$ TeV (dashed), $1\,$TeV (solid). The cyan overlay (dotted/dashed) denotes regions favored by low-energy $B$ physics. The constraints favor a combination of a light $V_0$ and a heavy $V_{1/2}$ with $m_{V_0} \approx 1$ TeV and $m_{V_{1/2}} \gtrsim 20$ TeV, depending on the scale of the trilinear couplings $\kappa_{V_i}$ and $h_V$. The constraint (\eqref{eq:DibosonConstraint}) with $r=0.28$ from $WW$, $ZZ$ and $Z\gamma$ limits is superimposed in red.}
\label{combined}
\end{figure}

In Fig. \ref{regionsb} we depict $\sigma_{\gamma\gamma}$ as a function of $\kappa_{V_i} m_\chi/m_{V_i}^2$. We fix the loop factor $A_1(\tau) \sim 7$, after explicitly verifying that $A_1(\tau)$ varies by 3\% in the relevant region of parameter space. The residual dependence on the masses from the loop function is hence small and is ignored. As $\kappa_{V_0}m_{\chi}/m_{V_0}^2,\kappa_{V_{1/2}}m_{\chi}/m_{V_{1/2}}^2$ increases, the corresponding diphoton cross section increases and the observed excess can be explained with, e.g., $\kappa_{V_0}m_{\chi} / m_{V_0}^2 \approx 1$ and $ \kappa_{V_{1/2}}m_{\chi}/m_{V_{1/2}}^2 < 0.8$. The shaded red areas denote the parameter space excluded by $WW, ZZ, Z\gamma$ searches, cf. Eq.~\eqref{eq:DibosonConstraint}. The darker shade applies in the case of same-sign $\kappa$ couplings resulting in the limits $\kappa_{V_0}m_{\chi}/m_{V_0}^2 \gtrsim 0.2$ and $\kappa_{V_{1/2}}m_{\chi}/m_{V_{1/2}}^2 \lesssim 1.2$. The more constraining case of opposite-sign $\kappa$ couplings is depicted in light red giving the limits $\kappa_{V_0}m_{\chi}/m_{V_0}^2 \gtrsim 0.7$ and $\kappa_{V_{1/2}}m_{\chi}/m_{V_{1/2}}^2 \lesssim 0.4$

In Fig.~\ref{lines} we show the behavior of $\sigma_{\gamma\gamma}$ in terms of $m_{V_0}$ for different choices of $m_{V_{1/2}}$ with $\kappa_{V_i}=\frac{4}{3}m_{V_i}$. The diphoton cross section decreases with large leptoquark masses and a mass $m_{V_0} \approx 1$~TeV is preferred in good agreement with the input from rare $B$ decays. For a given value of $m_{V_{1/2}}$, the diphoton cross section requirement yields an upper bound on $m_{V_0}$. In the case $m_{V_{1/2}} = 50$~TeV, $m_{V_0} > 1.1$~TeV yields a too low diphoton cross section, while $m_{V_{1/2}} = 5$ TeV requires $m_{V_0} < 1.5$ TeV.

The benchmark point $m_{V_0} = 1$ TeV, $m_{V_{1/2}}=30$ TeV, $\kappa_{V_i}=\frac{4}{3}m_{V_i}$ yields $\sigma \approx 4.0$~fb in good agreement with Eq. (\ref{sigmadiph}). Intriguingly, the combined results of neutrino mass generation and the 750~GeV diphoton excess point to a similar region in the parameter space of leptoquark masses. As shown in Fig. \ref{combined} the overlay of all constraints points at a light leptoquark $m_{V_0} \approx 1$~TeV together with a heavy $m_{V_{1/2}} \gtrsim 20$~TeV, depending on the size of the trilinear couplings $\kappa_{V_i}$ and $h_V$.
\section{Conclusion \label{seccon}}

The same leptoquarks that explain the rare $B$ decay anomalies can not only generate naturally small Majorana neutrino masses but also produce a large diphoton cross section to account for the recently observed $750$ GeV excess.

By proposing a simple framework based on an FN mechanism, we have shown that addressing several issues at the same time is entirely feasible and need not be overly fine-tuned. The total additional field content necessary to explain $R_K$, $R_D$, the 750 GeV excess and neutrino masses and mixing, includes no more than two vector leptoquarks and two SM singlet scalars. Only one additional symmetry is required to shape the fermion mass matrices and leptoquark couplings to comply with experimental data. We should note though that the model setup cannot be considered complete as we do not discuss the mechanism of mass generation for the vector leptoquarks. This could for example be accomplished through the breaking of a larger gauge group under which the vector leptoquarks are charged or by interpreting them as composite states \cite{Hernandez:2011rw}.

Our analysis shows that two leptoquarks with masses $m_{V_0} \approx 1 $TeV and $m_{V_{1/2}} \gtrsim 30$ TeV are favored to explain the diphoton excess and the lightness of neutrino masses. Furthermore, the model predicts dominant third generation leptoquark decays, mostly into $b\tau$ final states, and an enhanced $D^+ \rightarrow \pi^+ \nu\nu$ rate for indirect leptoquark searches. On the other hand, the already tightly constrained inclusive decay $B \rightarrow X_s \nu\nu$ remains SM like. 
\section{Acknowledgment}
We thank A. Crivellin and W.C. Huang for helpful discussions. H.P. is supported by DFG Grant No. PA 803/10-1. S.K. is supported by the `New Frontiers' program of the Austrian Academy of Sciences.

\end{document}